\documentclass[aps,prc,twocolumn,showpacs,nofootinbib,superscriptaddress]{revtex4} %  floatfix,,nofootinbib
\usepackage{graphicx,longtable} %,epsfig,
\usepackage{dcolumn}
\usepackage{mathptmx}
\usepackage{amsmath}
\usepackage[pdftex]{hyperref}
\usepackage{wasysym}
\usepackage[sort&compress]{natbib}

\newcommand{\beqy}{\begin{eqnarray}}
\newcommand{\eeqy}{\end{eqnarray}}
\newcommand{\bmlet}{\begin{subequations}}
\newcommand{\emlet}{\end{subequations}}
\newcounter{saveeqn}

\def\gsimeq{\,\,\raise0.14em\hbox{$>$}\kern-0.76em\lower0.28em\hbox  
{$\sim$}\,\,}  
\def\lsimeq{\,\,\raise0.14em\hbox{$<$}\kern-0.76em\lower0.28em\hbox  
{$\sim$}\,\,}  

\begin{document}

\title{Transitional $\gamma$ strength in Cd isotopes}

\author{A.~C.~Larsen}
\email{a.c.larsen@fys.uio.no}
\affiliation{Department of Physics, University of Oslo, N-0316 Oslo, Norway}
\author{I.~E.~Ruud}
\affiliation{Department of Physics, University of Oslo, N-0316 Oslo, Norway}
\author{A.~B\"{u}rger}
\affiliation{Department of Physics, University of Oslo, N-0316 Oslo, Norway}
\author{S.~Goriely}
\affiliation{Institut d'Astronomie et d'Astrophysique, 
    Universit\'e Libre de Bruxelles, CP 226,  1050 Brussels, Belgium}
%\email{sgoriely@astro.ulb.ac.be}
\author{M.~Guttormsen}
\affiliation{Department of Physics, University of Oslo, N-0316 Oslo, Norway}
\author{A.~G\"{o}rgen}
\affiliation{Department of Physics, University of Oslo, N-0316 Oslo, Norway}
\author{T.~W.~Hagen}
\affiliation{Department of Physics, University of Oslo, N-0316 Oslo, Norway}
\author{S.~Harissopulos}
\affiliation{Institute of Nuclear Physics, NCSR "Demokritos", 153.10 Aghia Paraskevi, Athens, Greece}
\author{H.~T.~Nyhus}
\affiliation{Department of Physics, University of Oslo, N-0316 Oslo, Norway}
\author{T.~Renstr{\o}m}
\affiliation{Department of Physics, University of Oslo, N-0316 Oslo, Norway}
\author{A.~Schiller}
\affiliation{Department of Physics and Astronomy, Ohio University, Athens, Ohio 45701, USA}
\author{S.~Siem}
\affiliation{Department of Physics, University of Oslo, N-0316 Oslo, Norway}
\author{G.~M.~Tveten}
\affiliation{Department of Physics, University of Oslo, N-0316 Oslo, Norway}
\author{A.~Voinov}
\affiliation{Department of Physics and Astronomy, Ohio University, Athens, Ohio 45701, USA}
\author{M.~Wiedeking}
\affiliation{iThemba LABS, P.O.  Box 722, 7129 Somerset West, South Africa}

\date{\today}

\begin{abstract}
The level densities and $\gamma$-ray strength functions of $^{105,106,111,112}$Cd have been extracted 
from particle-$\gamma$ coincidence data using the Oslo method. The level densities are in very 
good agreement with known levels at low excitation energy. The $\gamma$-ray strength functions display
no strong enhancement for low $\gamma$ energies. However, more low-energy strength is apparent for 
$^{105,106}$Cd than for $^{111,112}$Cd. For $\gamma$ energies
above $\approx$ 4 MeV, there is evidence for some extra strength, similar to what has been previously observed 
for the Sn isotopes. The origin of this extra strength is unclear; it might be due to $E1$ and $M1$ transitions 
originating from neutron skin oscillations or the spin-flip resonance, respectively. 

\end{abstract}

% PACS Numbers: 21.10.Ma (level density), 
%               21.10.-k (Properties of nuclei; nuclear energy levels), 
%               21.60.Jz (HF & QRPA)
% 		25.20.Lj (Photoproduction reactions)
%		24.30.Gd Other resonances
%		25.55.Hp	3He transfer reactions
%		27.40.+z	39 ² A ² 58
%		27.50.+e	59 ² A ² 89
%		27.60.+j	90 ² A ² 149
%		25.40.Hs	Transfer reactions, nucleon-induced
% From Voinov's PRL: 25.40.Lw, 25.20.Lj, 25.55.Hp, 27.40.+z

\maketitle

\section{Introduction}
\label{sec:int}

Recent measurements on the $\gamma$-strength function of several nuclei in the Fe-Mo mass region have revealed
an unexpected enhancement for low $\gamma$ energies ($E_\gamma \leq 3-4$ MeV)~\cite{Fe_Alex,Mo_RSF,V,Sc,magne_46Ti}. 
However, no such feature was seen in the heavier Sn isotopes~\cite{Sn_RSF,Heidi_Sn2} or in the rare-earth 
region~\cite{undraa,bagheri,hildes_DyRSF}.

For $^{95}$Mo, this low-energy enhancement has very recently been confirmed by 
an independent measurement and method~\cite{wiedeking95Mo}. It has also been shown in Ref.~\cite{larsen_goriely},
that if this increase persists in exotic nuclei close to the neutron drip line, it could boost  
the Maxwellian-averaged neutron-capture cross sections up to two orders of magnitude. 

However, as of today, there are more questions than answers regarding the low-energy enhancement. There is no
theoretical work predicting such a behavior, the underlying physics is unknown, neither the multipolarity nor
the electromagnetic character have been determined, and nobody knows for which nuclei the onset of this 
structure takes place. 

So far, there is only one nucleus, $^{60}$Ni, where there are strong indications that the enhancement
is due to $M1$ transitions~\cite{Ni_Alex}. One should however be careful to draw any general conclusions, 
because $^{60}$Ni is in many ways a special case. It has only positive-parity states
below excitation energies of $\approx 4.5$ MeV, which has significant consequences for the two-step cascade method 
employed in Ref.~\cite{Ni_Alex}. As discussed in Ref.~\cite{Ni_Alex}, it means that for the secondary $\gamma$ ray, 
$M1$ transitions are strongly enhanced compared to $E1$ transitions. 

The motivation for this work is to determine the transitional region of the low-energy enhancementÊ 
by investigating the $\gamma$-strength function of Cd isotopes using the Oslo method. 
The Cd isotopes have $Z=48$ and are in between 
Sn ($Z=50$) and Mo ($Z=42$). Thus, these experiments are a part of the experimental campaign exploring 
the onset of the low-energy enhancement.

In Sec.~\ref{sec:exp}, we give the experimental details and briefly describe the data analysis. 
In Sec.~\ref{sec:nor}, the normalization procedure of the level densities and $\gamma$-strength functions
is discussed. Further, we compare the measured $\gamma$-strength functions with semi-empirical models
in Sec.~\ref{sec:dis}. Finally, we give a summary and outlook in Sec.~\ref{sec:sum}.

\section{Experimental details and data analysis}
\label{sec:exp}

The experiments were performed at the Oslo Cyclotron Laboratory (OCL), utilizing a 38-MeV $^{3}$He 
beam delivered by the Scanditronix cyclotron. In the first experiment, the beam was bombarding 
a self-supporting target of $^{106}$Cd (96.7\% enrichment) with mass thickness
1.1 mg/cm$^2$. Typical beam currents were $0.3-0.5$ electrical nA (charge state $^3$He$^{2+}$). 
In the second experiment, the target was 99.5\% $^{112}$Cd with mass thickness $0.95$ 
mg/cm$^2$.  The beam current was $\approx 0.1-0.2$ electrical nA (charge state $^3$He$^{2+}$). 
Both experiments were run for five days. 
The reactions of interest are  $^{106,112}$Cd($^3$He,$^3$He$^\prime$$\gamma$)$^{106,112}$Cd and
$^{106,112}$Cd($^3$He,$\alpha\gamma$)$^{105,111}$Cd. The
$Q$-values of the pick-up reactions are $9703.9(124)$ keV 
and $11183.295(3)$ keV, respectively~\cite{qcalc}. 

Particle-$\gamma$ coincidences were measured
with the Silicon Ring (SiRi) particle-detector system~\cite{siri} 
and the CACTUS array for detecting $\gamma$ rays~\cite{CACTUS}. 
The SiRi system consists of eight 130-$\mu$m thick silicon detectors, where each of them 
is divided into eight strips. One strip has an angular resolution of $\Delta\theta = 2^{\circ}$.
Each of these segmented, thin detectors are put in front of a 1550-$\mu$m thick back detector. 
The full SiRi system has then 64 individual detectors in total, covering scattering angles between 
$40-54^{\circ}$ and a solid-angle coverage of $\approx 6$\%. For the Cd experiments, SiRi was placed
in forward angles with respect to the beam direction. 
 
The CACTUS array consists of 28 collimated $5" \times 5"$ NaI(Tl) crystals. 
The total efficiency of CACTUS is $15.2(1)$\% at $E_\gamma = 1332.5$ keV. 
%Also a 65\% HPGe detector was placed in the CACTUS frame to monitor the experiment.
The charged ejectiles and the 
$\gamma$-rays were measured in coincidence event-by-event, with time resolution of $\approx 15$ ns.

Using the $\Delta E-E$ technique, each charged-particle species was identified. Gates were set
on the $^3$He and $\alpha$ ejectiles to select the correct reaction channel. 
Furthermore, the reaction kinematics and the known $Q$-value for the reaction allowed us to relate the measured
ejectile energy to the excitation energy of the residual nucleus.  

\begin{figure}[bt]
 \begin{center}
 \includegraphics[clip,width=\columnwidth]{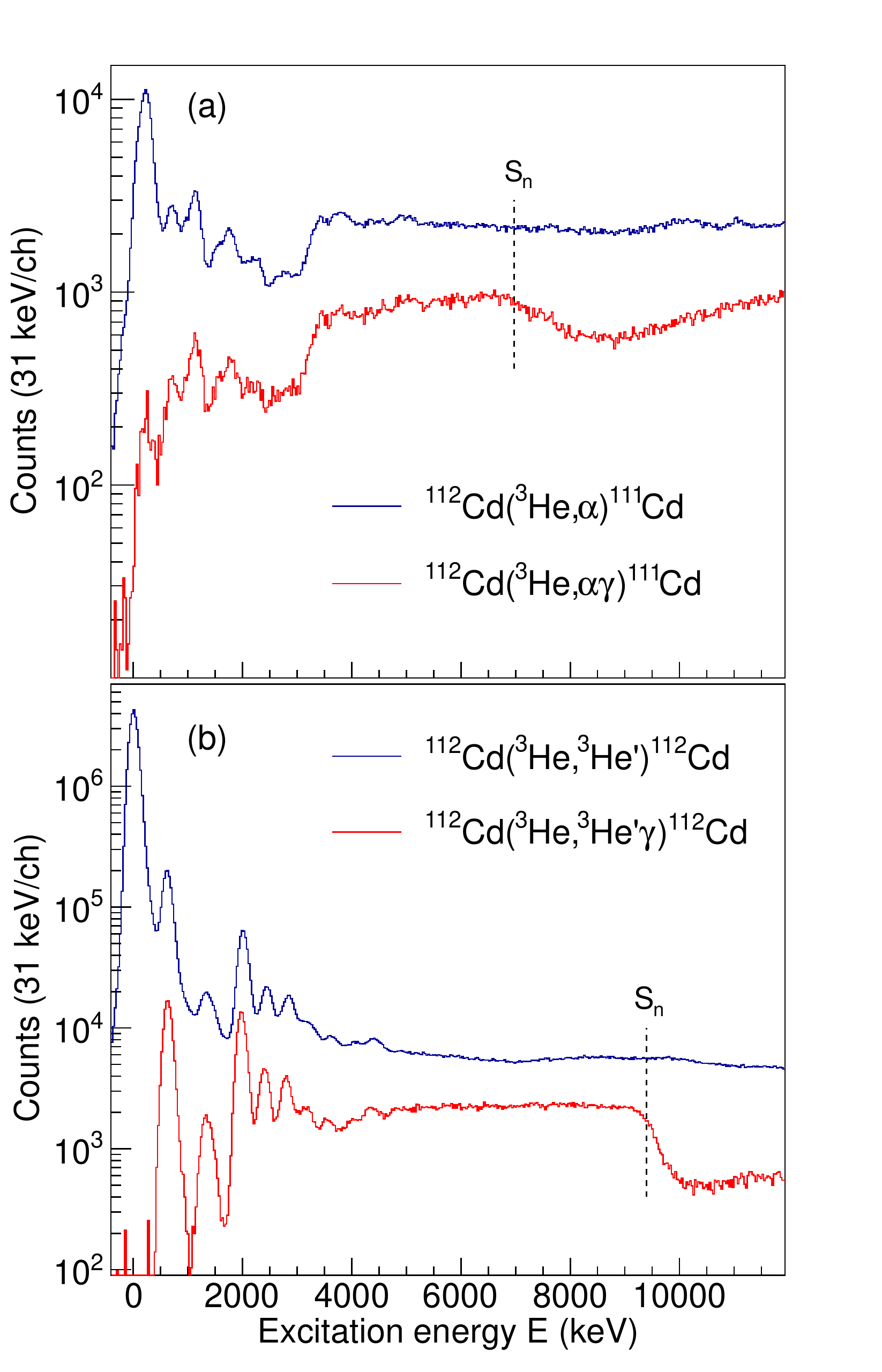}
%\vskip 2cm
 \caption{(Color online) Singles particle spectra (blue) and in coincidence
 with $\gamma$ rays (red) from (a) the $^{112}$Cd($^3$He,$\alpha)$ reaction and
 (b) the $^{112}$Cd($^3$He,$^3$He$^\prime$) reaction. 
 The dashed lines indicate the neutron separation energies for the final nucleus.}
 \label{fig:particlespectra}
 \end{center}
 \end{figure}

In Fig.~\ref{fig:particlespectra}, the $^3$He and $\alpha$ spectra with and without $\gamma$-coincidence
requirements are shown. 
%---------------------------------------------------%
%---------------------------------------------------%
It is interesting to see how the $^3$He and $\alpha$ spectra in coincidence with 
$\gamma$ rays differ at the neutron separation energy. They both display a drop because the 
neutron channel is open. However, while the $^3$He spectrum shows a rather abrupt drop 
(compatible  with the energy resolution of $\approx 200$ keV), the slope of the $\alpha$ spectrum 
is much less steep and a minimum is not reached until $\approx S_n+1.5$ MeV. This can be explained by 
considering the final nuclei in the reactions $^{112}$Cd($^3$He,$^3$He$^\prime$$n\gamma$)$^{111}$Cd and
$^{112}$Cd($^3$He,$\alpha n\gamma$)$^{110}$Cd. In the latter case, the odd, final nucleus $^{111}$Cd
has many states within a relatively broad spin window at low excitation energy. 
However, this is not so for $^{110}$Cd, where there are only $0^{+}$ and $2^{+}$ states below 
$\approx 1.5$ MeV. As the ($^3$He,$\alpha$) reaction favors high-$\ell$ transfer in general, the populated 
states very likely have an average spin larger than $2$. Thus,
there is an effective spin hindrance which explains the observed behavior in the $\alpha$ spectrum.

The $\gamma$-ray spectra for each excitation-energy bin 
were unfolded using the known response functions of the CACTUS array, as 
described in Ref.~\cite{gutt6}. The main advantage of this method
is that the experimental statistical uncertainties are preserved, without introducing new, artificial
fluctuations.

\begin{figure}[bt]
 \begin{center}
 \includegraphics[clip,width=\columnwidth]{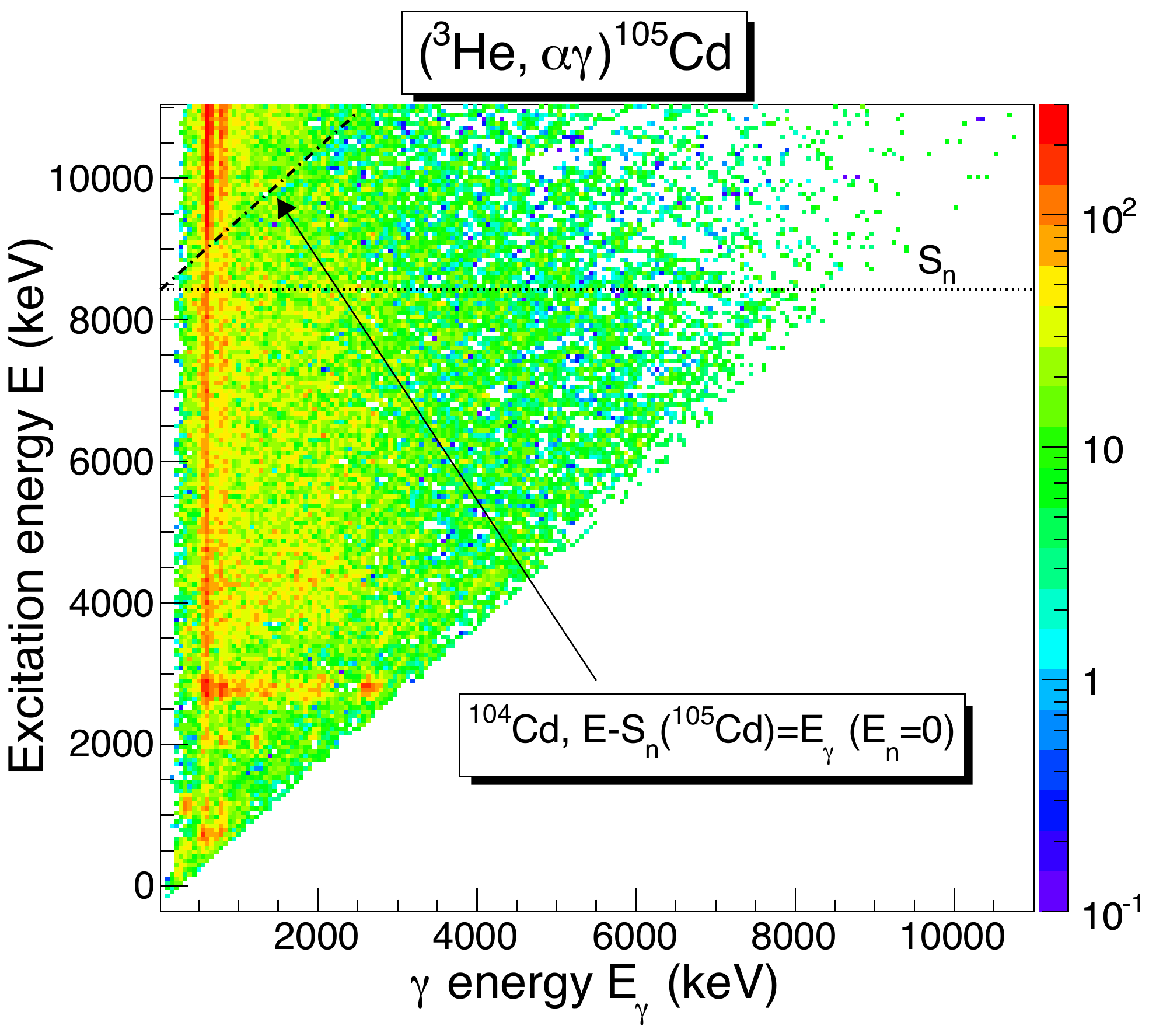}
 \caption{(Color online) Excitation energy vs. $\gamma$ energy matrix for $^{105}$Cd.
 The $\gamma$-ray spectra are unfolded for each excitation-energy 
 bin. The dashed line indicates the neutron separation energy in $^{105}$Cd.
 The dashed-dotted line shows where the $E=E_{\gamma}$ diagonal would be in $^{104}$Cd
 for the extreme case where the outgoing neutron has zero kinetic energy. }
 \label{fig:alfnaun}
 \end{center}
 \end{figure}

The matrix of unfolded $\gamma$ spectra for each excitation-energy bin is shown 
for $^{105}$Cd in Fig.~\ref{fig:alfnaun}. One may notice a peculiar feature in this matrix. 
Surprisingly, there is a considerable amount of $\gamma$ rays from $^{105}$Cd that survive 
several MeV above $S_{n}$, see the region to the right of the dashed-dotted line in 
Fig.~\ref{fig:alfnaun}. For example, the intensity of 5-MeV $\gamma$ rays is practically the same 
for the excitation-energy region $7.0-8.0$ MeV and $8.5-9.5$ MeV. This could be caused by the 
difference in spin between the populated initial states 
and the spin of the first excited states in $^{104}$Cd ($2^+,4^+$).

%---------------------------------------------------%
%---------------------------------------------------%

After the $\gamma$ spectra were unfolded, the distribution of first-generation $\gamma$ 
rays\footnote{The first $\gamma$ ray emitted in the decay cascade.} for 
each excitation-energy bin was extracted via an iterative subtraction technique~\cite{Gut87}. 
The basic assumption of this method is that the decay routes are the same regardless of the 
population mechanism of the initial states (either directly via the nuclear reaction or
from $\gamma$ decay from above-lying states). For a discussion of uncertainties and possible errors
of the first-generation method, see Ref.~\cite{systematic}.   

From the excitation energy vs. first-generation $\gamma$-ray matrix, one can extract the
functional form of the level density and the $\gamma$ transmission coefficient. This is done
with an iterative procedure as described in Ref.~\cite{Schiller00}, with the following ansatz:
\begin{equation}
P(E, E_{\gamma}) \propto  \rho (E_{\mathrm{f}}) {\mathcal{T}}  (E_{\gamma}).
\label{eq:brink}
\end{equation}
Here, $P(E,E_{\gamma})$ is the experimental first-generation matrix, $\rho(E_{\mathrm{f}})$ is
the level density at the final excitation energy $E_{\mathrm{f}}$, with 
$E_{\mathrm{f}} = E - E_{\gamma}$, and ${\mathcal{T}}(E_{\gamma})$ is the $\gamma$-transmission
coefficient. Every point of the $\rho$ and ${\mathcal{T}}$ functions is assumed to be an independent 
variable, and a global $\chi^{2}$ minimum is reached typically within 10--20 iterations.

The method is based on the assumption that 
the reaction leaves the product nucleus in a compound state, which then subsequently decays in a manner
that is independent on the way it was formed, i.e. a statistical decay process~\cite{BM}. Therefore,
a lower limit is set in the excitation energy to ensure that decay from compound states dominates 
the spectra. In addition, an upper excitation-energy limit at $\approx S_{n}$ is 
employed\footnote{When the neutron channel is open, the excitation energy is not well defined anymore,
because neutron energies are not measured.}. Because of methodical problems with the first-generation method 
for low $\gamma$ energies, $\gamma$ rays below $\approx 1.0$ and $1.5$ MeV for $^{105,111}$Cd and 
$^{106,112}$Cd, respectively, were excluded from the further analysis (see also Ref.~\cite{systematic}).

The $\gamma$-transmission coefficient ${\cal T}$ is a function of $E_\gamma$ only, in accordance with
the Brink hypothesis~\cite{brink}, which in its generalized form states that any collective decay 
mode has the same properties whether it is built on the ground state or on excited states. 
This assumption is proven to be incorrect for nuclear reactions involving high temperatures and/or spins, 
see for example Ref.~\cite{Andreas&Thoennessen}. However, in the present work, neither high-spin states nor
high temperatures are reached ($T_{\mathrm{f}} \propto \sqrt{E_{\mathrm{f}}}$, and the populated spin 
range is centered within $J \sim 2 - 8 \hbar$). Therefore, eventual spin and/or temperature dependencies
should not have a significant impact on the results.

\section{Normalization of level density and $\gamma$-strength function}
\label{sec:nor}

The extracted level density and the $\gamma$-ray transmission coefficient
give identical fits to the experimental
data with the transformations~\cite{Schiller00} 
\begin{eqnarray}
\label{eq:array1}
\tilde{\rho}(E-E_\gamma)&=&{\mathcal{A}}\exp[\alpha(E-E_\gamma)]\,\rho(E-E_\gamma),\\
\tilde{{\mathcal{T}}}(E_\gamma)&=&{\mathcal{B}}\exp(\alpha E_\gamma){\mathcal{T}} (E_\gamma).
\label{eq:array2}
\end{eqnarray}
Therefore, 
the transformation parameters ${\mathcal{A}}$, $\alpha$, and ${\mathcal{B}}$ were determined from external data. 

\subsection{Level density}
\label{subsec:nld}
For the level density, the absolute normalization ${\mathcal{A}}$ and the slope $\alpha$ can be 
determined from the known, discrete levels~\cite{ENSDF} at low excitation energy, and from
neutron-resonance spacings at the neutron separation energy $S_n$~\cite{RIPL}. For the latter,
we must estimate the total level density at $S_n$ from the neutron resonances, which are for 
a few spins only. Also, because of the selected lower limit of $E_\gamma$ for the extraction of 
$\rho$ and ${\mathcal{T}}$ (see Sec.~\ref{sec:exp}), our level-density data reach up to $E \approx S_n - 1$ MeV.
Therefore, we must interpolate between our data and the level density at $S_n$. We have here chosen to 
use the back-shifted Fermi gas (FG) model with the parameterization of von Egidy and Bucurescu~\cite{egidy2}
for that purpose.

Because the spin distribution is poorly known at high excitation energies, 
a systematic uncertainty will be introduced to the slope of the level density and $\gamma$-strength function
(see Ref.~\cite{systematic} for a thorough discussion on this subject). In addition, the light-ion
reactions in the experiments populate only a certain spin range, which usually is for rather low spins. 
Therefore, the full spin distribution should also be folded with the experimental
spin distribution. 

In this work, we have 
tested two different approaches to normalize the level densities. First, we have used 
the back-shifted Fermi gas parameterization of von Egidy and Bucurescu~\cite{egidy2} to 
estimate the total level density at the neutron separation energy,
$\rho(S_n)$. Second, we have used
the microscopic level densities of Goriely, Hilaire and Koning~\cite{go08} at high excitation energies. These 
level densities are calculated within the combinatorial plus Hartree-Fock-Bogoliubov
approach, and are resolved in spin and parity. The applied parameters are listed in Tab.~\ref{tab:nldpar},
together with the Fermi-gas parameters of Ref.~\cite{egidy2} used for the interpolation between our
data and the estimated $\rho(S_n)$.

We start with the back-shifted Fermi gas approach. 
We adopt the expression for the spin cutoff parameter from Ref.~\cite{egidy2}:
\begin{equation}
\sigma^2(E) = 0.0146 A^{5/3} \frac{1+\sqrt{1+4a(E-E_1)}}{2a},
\label{eq:spincut}
\end{equation}
where $A$ is the mass number, $a$ is the level density parameter and $E_1$ is the backshift parameter
(see Ref.~\cite{egidy2} for further details).
The total level density can be calculated by
\begin{equation}
\rho(S_n) = \frac{2\sigma^2}{D_0} \cdot 
    \frac{1}{(I_t+1)\exp\left[-(I_t+1)^2/2\sigma^2\right] + I_t\exp\left[-I_t^2/2\sigma^2\right]},
\label{eq:oldD}
\end{equation} 
where $D_0$ is the level spacing of $s$-wave neutrons and $I_t$ is the ground-state spin of the
target nucleus in the $(n,\gamma)$ reaction. In Eq.~(\ref{eq:oldD}), it is
assumed that both parities contribute equally to the level density at $S_n$ (see Refs.~\cite{Schiller00}
and~\cite{systematic}).

From the Fermi-gas calculation, we get $\rho_{\mathrm{FG}}(S_n)$, which differs somewhat from the 
semi-experimental value $\rho(S_n)$. Therefore, a correction factor $\eta$ is applied to ensure that 
the Fermi-gas interpolation matches $\rho(S_n)$ (see Tab.~\ref{tab:nldpar}).

%************************************************************************************%
\begin{table*}[htb]
\caption{Parameters used for the calculation of $\rho(S_n)$ (see text).} 
\begin{tabular}{lccccccccccccc}
\hline
\hline
Nucleus    & $I_t^\pi$  & $D_0$        & $S_n$ & $\sigma(S_n)$ & $a$          & $E_1$  & $\rho_{\mathrm{FG}}(S_n)$ & $\rho(S_n)$        & $\eta$ & $\tilde{\sigma}(S_n)$ & $\tilde{\rho}(S_n)$ & shift $E_{\mathrm{HFB}}$  & range $I_i$   \\
		   &            & (eV)         & (MeV) &               & (MeV$^{-1}$) & (MeV)  &  ($10^5$ MeV$^{-1}$)      & ($10^5$ MeV$^{-1}$)&        &                       & ($10^5$ MeV$^{-1}$) &(MeV)                     & ($\hbar$)      \\
\hline
$^{105}$Cd &  $0^+$     & $-$ & 8.427 & 5.71          & 10.88        & -0.567 &  1.43                     & 1.78(89)$^a$       & 1.25   &    4.5                &   1.11(56)$^a$      & 0.042                     & $1/2 - 13/2$  \\
$^{106}$Cd &  $5/2^+$   & $-$ &10.874 & 5.85          & 11.39        &  0.746 &  6.44                     & 8.05(40)$^a$       & 1.25   &    4.5                &   5.3(26)$^a$       & 0.052                     & $0 - 6$       \\
$^{111}$Cd &  $0^+$     & 155(20)      & 6.976 & 5.43          & 13.56        & -0.640 &  2.99                     & 3.87(91)           & 1.29   &    4.5                &   2.68(72)          & 0.435                     & $1/2 - 13/2$  \\
$^{112}$Cd &  $1/2^+$   & 27(2)        & 9.394 & 5.61          & 13.82        &  0.713 & 11.9                      & 12.0(25)           & 1.01   &    4.5                &   7.8(16)           & 0.540                     & $0 - 6$       \\
\hline
\hline
\end{tabular}
\\
\label{tab:nldpar}
$^a$ Estimated from systematics.
\end{table*}
%************************************************************************************%

\begin{figure}[bt]
 \begin{center}
 \includegraphics[clip,width=\columnwidth]{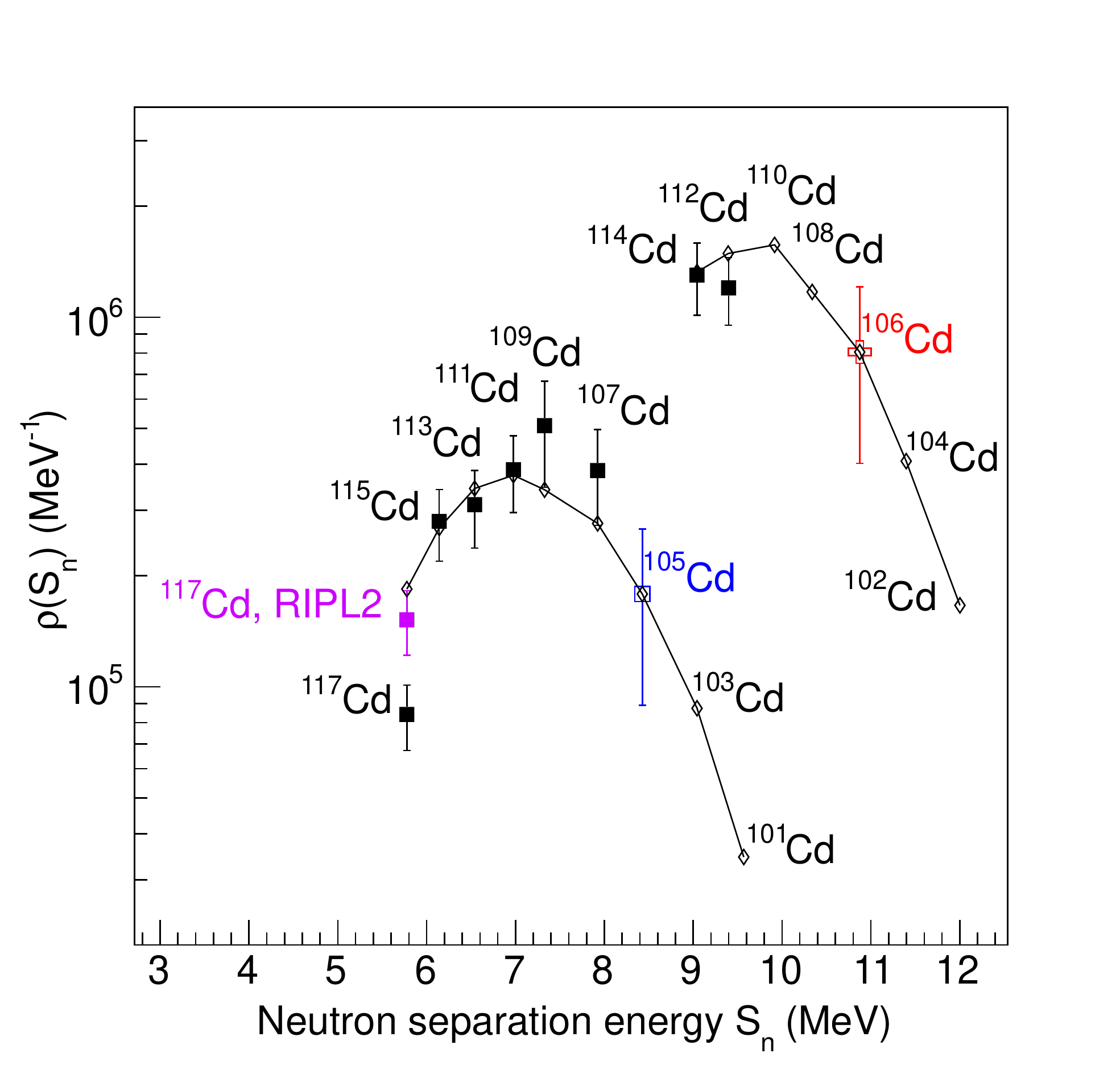}
 \caption {(Color online) Estimation of $\rho(S_n)$ for $^{105,106}$Cd. The filled, black squares are 
 calculated from known neutron resonance spacings in RIPL-3~\cite{RIPL} using Eq.~(\ref{eq:oldD}) with 
 $\sigma$ values from Ref.~\cite{egidy2}. 
 The filled, violet square is the result for $^{117}$Cd using the $D_0$ value 
 recommended in RIPL-2. The small, open diamonds connected with lines are calculated values from the 
 back-shifted Fermi gas approach~\cite{egidy2} multiplied 
 with a common factor of 1.25 to bring them within the error bars of the semi-experimental $\rho(S_n)$ values.
 The blue, open square and the red, open cross are the estimated values for $\rho(S_n)$ of $^{105,106}$Cd, respectively.}
 \label{fig:rhosyst}
 \end{center}
 \end{figure}

As there is no information on the level spacing for $^{105,106}$Cd ($^{104,105}$Cd are unstable), we have estimated 
the total level density at the neutron separation energy
from systematics for these nuclei, see Fig.~\ref{fig:rhosyst}. Here, we have calculated the 
semi-experimental $\rho(S_n)$ for all Cd isotopes where the neutron resonance spacing $D_0$ is known. 
For all $D_0$ values we have used the Reference Input Parameter Library (RIPL-3) evaluation~\cite{RIPL}, 
except for $^{117}$Cd where we have also used the RIPL-2 value. 
%---------------------------------------------------%
%---------------------------------------------------%

It is striking how the values of $\rho(S_n)$ actually decrease 
as a function of $S_n$ for the isotopes with $A\leq 108$. This is probably an effect of approaching the $N=50$ 
closed shell. It is, however, unfortunate that there are no experimental $D_0$ values for these nuclei, 
so the uncertainty of the estimated $\rho(S_n)$ for $^{105,106}$Cd must necessarily be large; 
we have assumed a 50\% uncertainty. 

\begin{figure}[bt]
 \begin{center}
 \includegraphics[clip,width=\columnwidth]{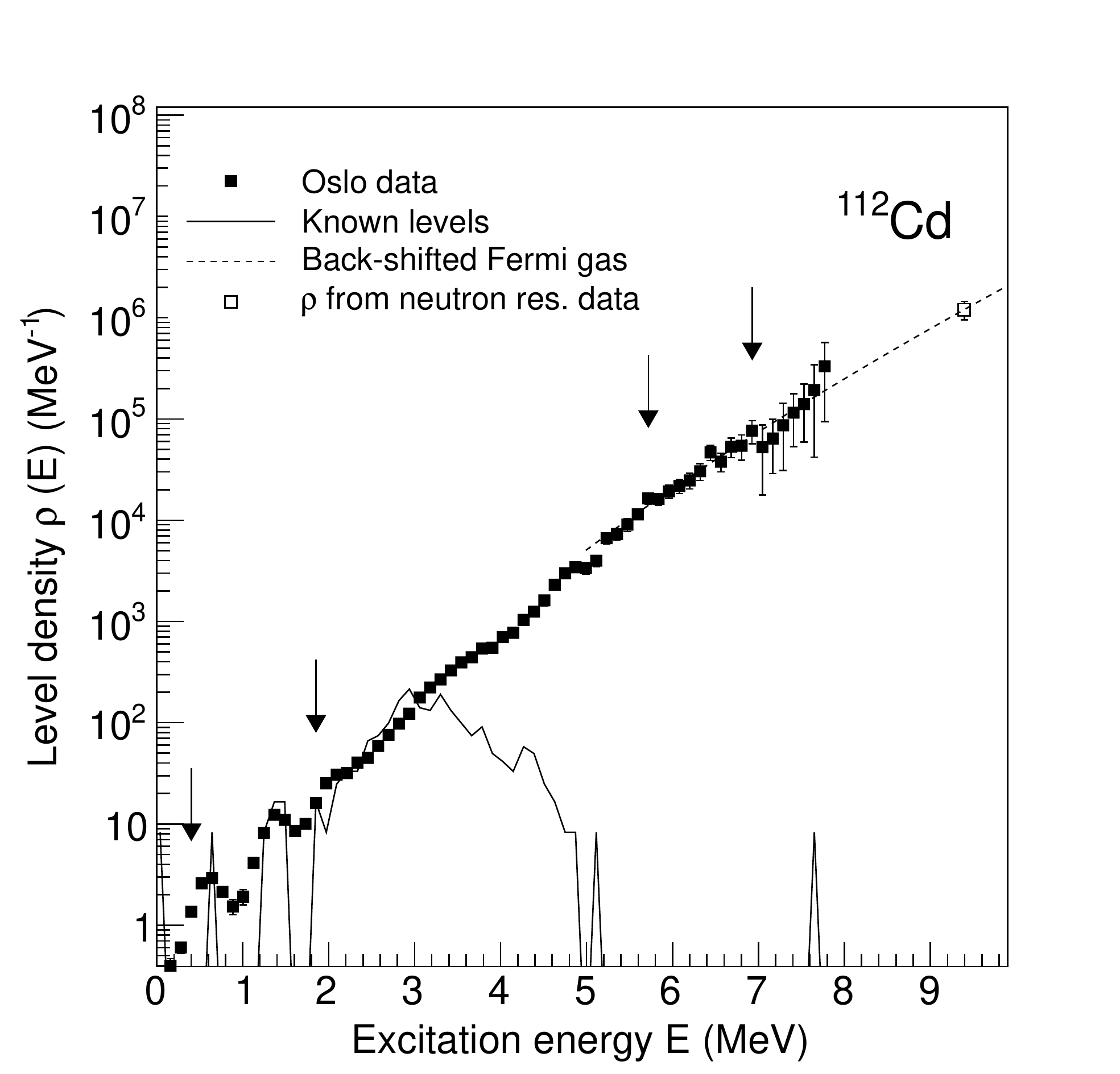}
 \caption {Normalization of the level density of $^{112}$Cd to the known, discrete levels (jagged line), 
    and $\rho(S_n)$ (see text). }
 \label{fig:rhonorm}
 \end{center}
 \end{figure}

The normalization procedure is demonstrated for $^{112}$Cd in Fig.~\ref{fig:rhonorm}. 
%---------------------------------------------------%
%---------------------------------------------------%
The agreement between our data and the discrete levels~\cite{ENSDF} is very satisfying. We notice however that 
the ground state seems to be underestimated; this is probably because there are very few direct decays to the 
ground state, most of the decay goes through the first $2^+$ state. We also see that the triplet of two-phonon 
vibrational states $0^+$, $2^+$, $4^+$, at about $E \approx 1.4$ MeV, is clearly seen in our level-density data, 
as well as the one-phonon first excited $2^+$ state at $0.62$ MeV (see, e.g., Ref.~\cite{garrett} for a discussion
on the vibrational nature of Cd isotopes). 

\begin{figure}[bt]
 \begin{center}
 \includegraphics[clip,width=\columnwidth]{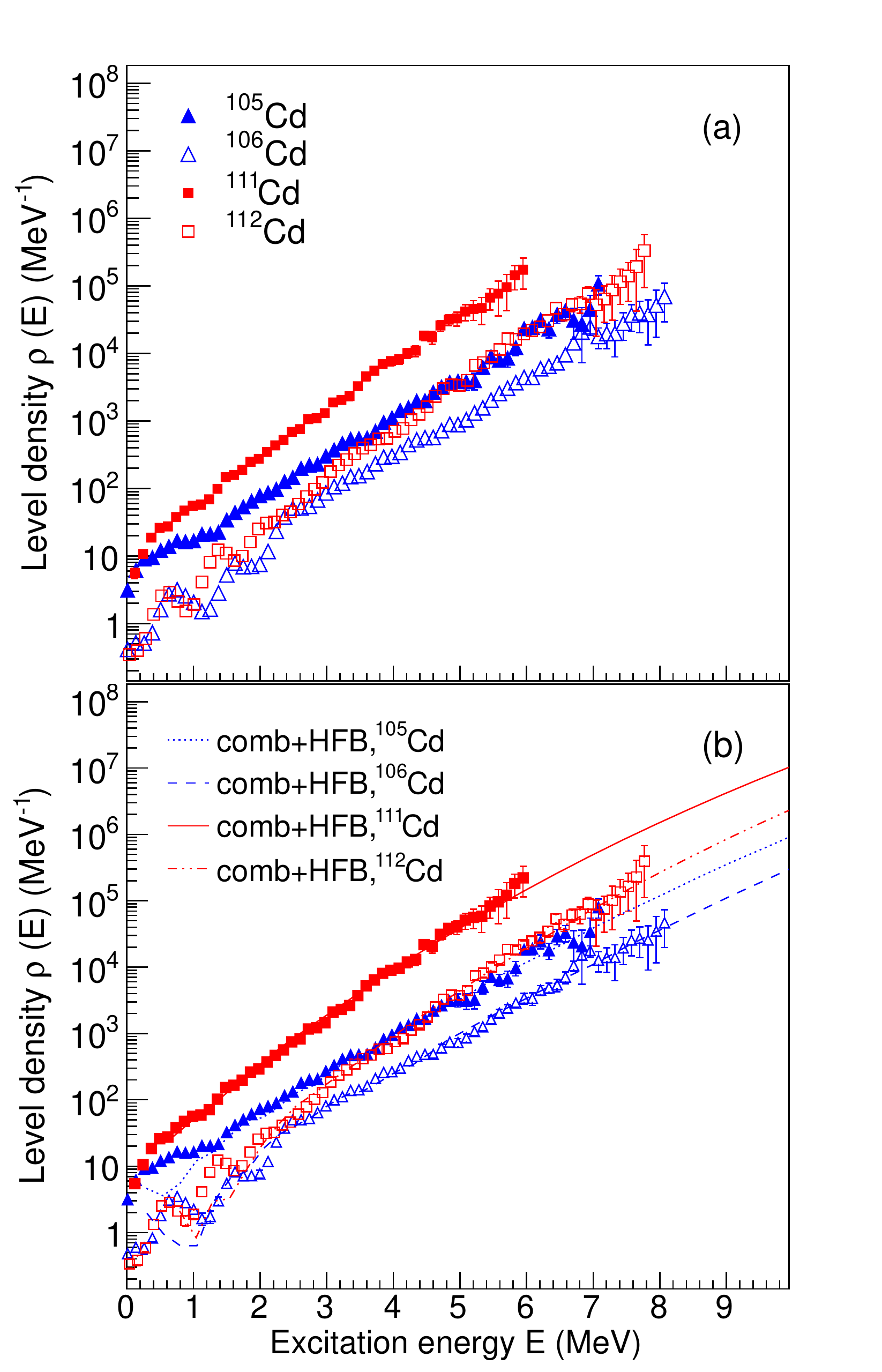}
 \caption {(Color online) Normalized level densities of $^{105,106,111,112}$Cd
    with (a) the fermi-gas approach and (b) the combinatorial plus Hartree-Fock-Bogoliubov approach.}
 \label{fig:rhoall}
 \end{center}
 \end{figure}

The level densities normalized with the back-shifted Fermi gas approach are shown in Fig.~\ref{fig:rhoall}a.
%---------------------------------------------------%
%---------------------------------------------------%
Again, the effect of approaching the $N=50$ closed shell is clearly seen. The slope in level density is smaller 
for $^{105,106}$Cd than for $^{111,112}$Cd. Also, we see that the level densities of the neighboring isotopes 
are parallel, but the increase in level density of the odd-$A$ nucleus compared to the even neighbor is 
smaller for $^{105}$Cd than for $^{111}$Cd. 

For the second approach, we have used the combinatorial plus HFB calculations of Ref.~\cite{go08}. 
Here, we have normalized our data to obtain a best fit to the microscopic level densities at high 
excitation energies ($E \geq 4-5$ MeV). As described
in Ref.~\cite{go08}, an energy shift is used in order to optimize the reproduction 
of the known, discrete levels. The applied energy shifts are listed in Tab.~\ref{tab:nldpar}. 

The level-density data normalized to the microscopic calculations are shown in Fig.~\ref{fig:rhoall}b.
It is seen that the two independent normalization methods yield very similar results. 

We have also taken into account that the spin distribution of the initial levels could be rather narrow.
As discussed in Ref.~\cite{guttormsen_spin}, the ($^3$He,$\alpha$) reaction in forward angles gives an 
average spin transfer of $\approx 5\hbar$ at $E\approx 5 $ MeV in the rare-earth region. For excitation 
energies below 3 MeV, it is shown in Ref.~\cite{dracoulis} that the $^{106}$Cd($^3$He,$\alpha$)$^{105}$Cd 
reaction involves $\ell=2,4,$ and 5. 

Turning to the inelastic scattering, where 
vibrational states are favored, we see from the $^{106,112}$Cd data below $E\approx 3$ MeV that levels 
with $I=2,3,4$ are strongly populated. For levels with higher spins the data are inconclusive, but it is clear 
that they are significantly less populated. We therefore estimate a reduced spin cutoff parameter, 
$\tilde{\sigma}$, to be $\approx 4.5$ for all the Cd nuclei studied here. 
This corresponds to a reduced level density at $S_n$, $\tilde{\rho}(S_n)$. For the microscopic 
level densities, which are spin-dependent, we filter out the levels within the approximate 
experimental spin range (see Tab.~\ref{tab:nldpar}). 

\begin{figure}[bt]
 \begin{center}
 \includegraphics[clip,width=\columnwidth]{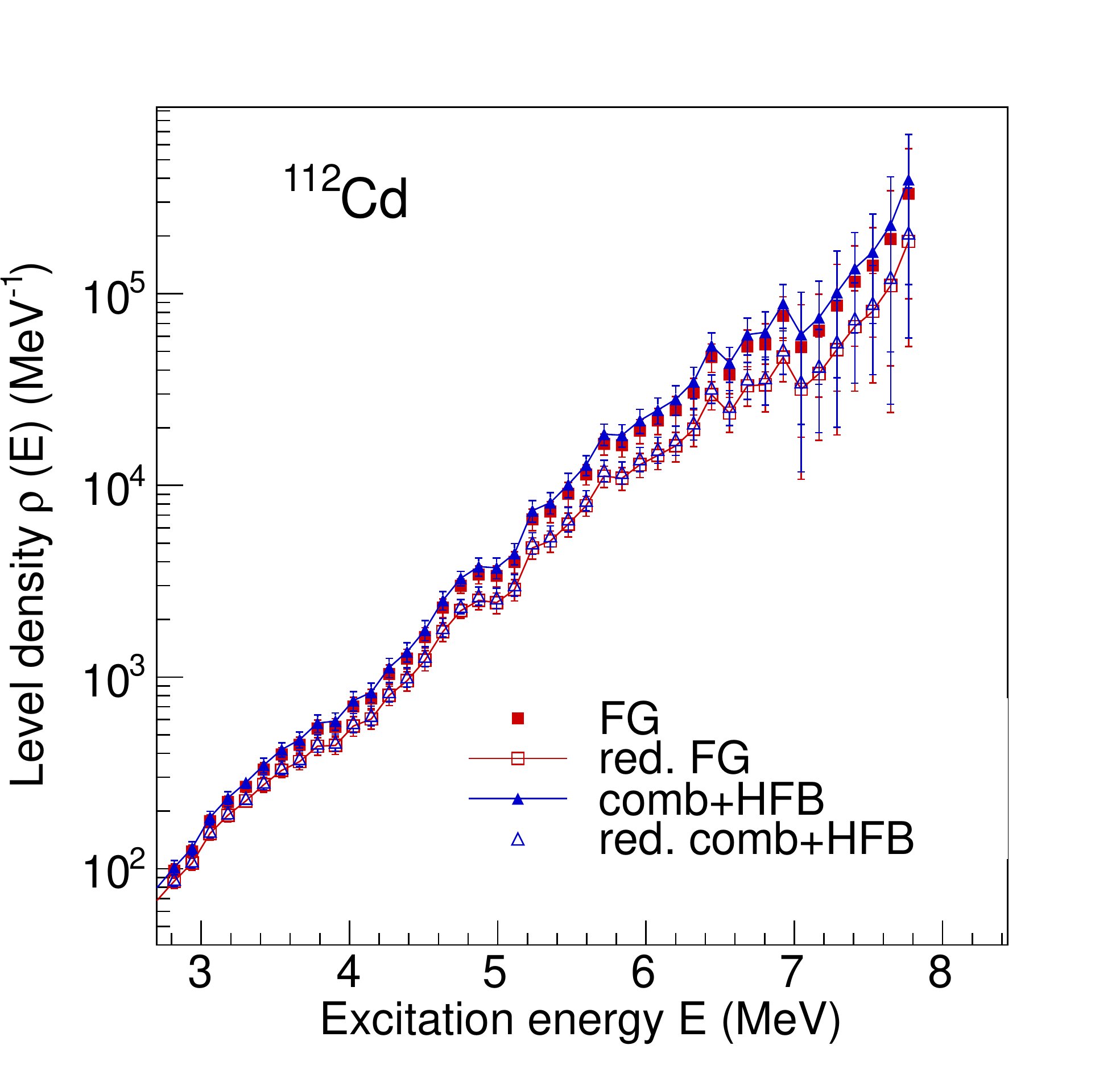}
 \caption {(Color online) The various normalizations of $^{112}$Cd: the fermi-gas approach (FG, red 
    squares), FG approach with a reduced spin-cutoff parameter (red, open squares), the combinatorial
    plus HFB approach (blue triangles) and with a reduced spin range (open, blue triangles). }
 \label{fig:rhored}
 \end{center}
 \end{figure}

The four different normalizations are shown for $^{112}$Cd for $E=3-8$ MeV in Fig.~\ref{fig:rhored}. 
%---------------------------------------------------%
%---------------------------------------------------%
As seen in this figure, the effect of the reduced spin range is not large at low excitation energies,
but could be as much as a factor of 2 for example at $E=7.9$ MeV.

\subsection{Gamma strength function}
\label{subsec:gsf}

The slope of the $\gamma$ strength function is given by the slope of the level density, 
see Eqs.~(\ref{eq:array1}) and~(\ref{eq:array2}). Therefore, the only parameter left
to determine is the absolute value $\mathcal{B}$. This is done using known values on the average, total radiative
width at $S_n$, $\left< \Gamma_{\gamma0} \right>$, extracted from $s$-wave neutron resonances~\cite{RIPL} 
by~\cite{voin1}: 
\begin{align}
\langle \Gamma_{\gamma}(S_n,I_t\pm 1/2,&\pi_t)\rangle =
 \frac{D_0}{4\pi}\int_{E_{\gamma}=0}^{S_n}\mathrm{d}E_{\gamma}{\mathcal{B}}\mathcal{T}(E_{\gamma}) \nonumber \\ 
 &\times \rho(S_n-E_{\gamma}) \sum_{I= -1}^{1} g(S_{n}-E_{\gamma},I_{t}\pm 1/2+I),
\label{eq:width}
\end{align}
where $I_t$ and $\pi_t$ are the spin and parity of the target nucleus in the $(n,\gamma)$ reaction, 
and $\rho(S_n-E_{\gamma})$ is the experimental level density. The spin distribution is assumed
to be given by~\cite{GC}:
\begin{equation}
g(E,I) \simeq \frac{2I+1}{2\sigma^2}\exp\left[-(I+1/2)^2/2\sigma^2\right]
\label{eq:spindist}
\end{equation}
for a specific excitation energy $E$, spin $I$, and a spin cutoff parameter $\sigma$.
All values are known in Eq.~(\ref{eq:width}) except the 
parameter $\mathcal{B}$, which can now be determined.

For $^{111,112}$Cd, the values for $\left< \Gamma_{\gamma0} \right>$ are 71(6) and 106(15) meV, respectively. 
However, again we lack neutron resonance data for $^{105,106}$Cd. We must therefore estimate  
$\left< \Gamma_{\gamma0} \right>$ and $D_0$ for these nuclei. For the FG approach, 
$D_0$ is evaluated from the previously
estimated $\rho(S_n)$ values (see Tab.~\ref{tab:nldpar}). We get $D_0 = 375(188)$ and 16.3(82) eV for
$^{105,106}$Cd, respectively. The combinatorial plus HFB calculations predict $D_0 = 294$ eV and 13.6 eV
for $^{105,106}$Cd, respectively. 
%---------------------------------------------------%
 \begin{figure}[bt]
 \begin{center}
 \includegraphics[clip,width=\columnwidth]{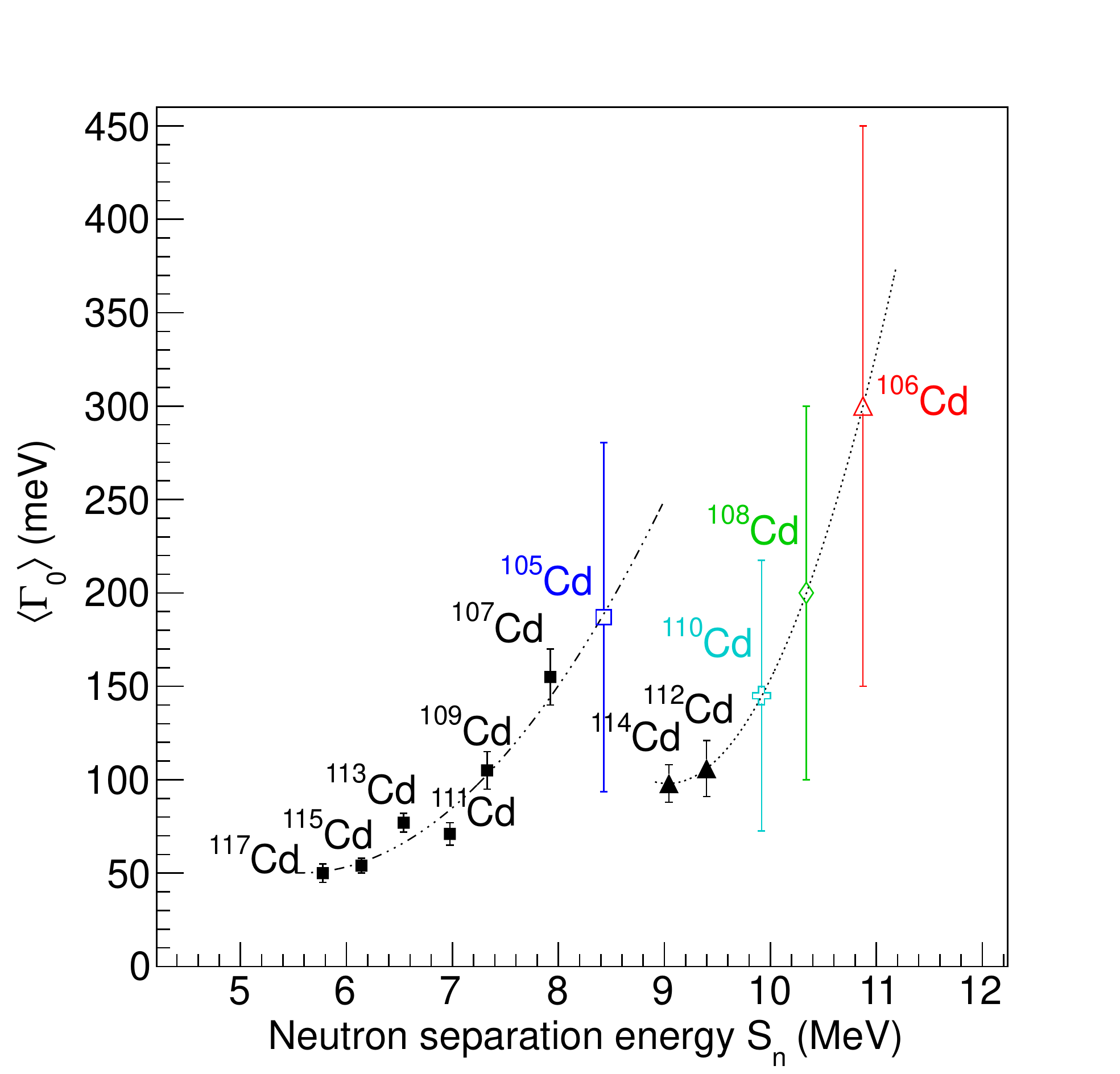}
 \caption {(Color online) Estimation of $\left< \Gamma_{\gamma0} \right>$ for $^{105,106}$Cd (see text). 
    The black squares are known values for odd Cd isotopes and the black triangles are for the even ones;
    all values are taken from Ref.~\cite{RIPL}. The dashed-dotted line represents the best fit with 
    a quadratic function for the odd nuclei. 
    The blue, open square is the estimated $\left< \Gamma_{\gamma} \right>$ for $^{105}$Cd, and the 
    red, open triangle for $^{106}$Cd. The dashed line indicates a quadratic function for the even isotopes
    in the same fashion as for the odd ones. Estimations of $^{108,110}$Cd are shown 
    for completeness (green, open
    diamond and cyan, open cross, respectively).}
 \label{fig:systG}
 \end{center}
 \end{figure}
%---------------------------------------------------%

To estimate the average total radiative width, we have considered systematics from the Cd isotopes
where $\left< \Gamma_{\gamma0} \right>$ is known, see Fig.~\ref{fig:systG}. 
It is difficult to predict with reasonable certainty 
the unknown values for $^{105,106}$Cd because of the possible shell effects. Because we also lack data on 
$^{108,110}$Cd, it is especially problematic for $^{106}$Cd. We have therefore also assumed that 
for $\gamma$ energies above $\approx 5-6$ MeV, the strength functions for all the Cd isotopes should be very
similar, because this region should be dominated by the low-energy tail of the Giant Electric Dipole Resonance (GDR).
The GDR is mainly governed by the number of protons, and thus it is reasonable to believe that the properties
should be the same for all Cd isotopes, at least to a large extent. 

As shown in Fig.~\ref{fig:systG}, we have fitted a quadratic function to the $\left< \Gamma_{\gamma0} \right>$ values 
of the odd Cd isotopes, 
and for $^{105}$Cd we estimate $\left< \Gamma_{\gamma0} \right> = 187(94)$ meV. For the
even isotopes, we only have two data points. However, considering the trend for the odd isotopes and claiming 
the postulated similarity of the strength functions at high $E_\gamma$, we have chosen a rather large 
value of 300(150) meV. To guide the eye, we have shown a quadratic fit as for the odd case, and displayed the 
predicted $\left< \Gamma_{\gamma0} \right>$ values also for $^{108,110}$Cd (see Fig.~\ref{fig:systG}).

\begin{figure*}[bt]
 \begin{center}
 \includegraphics[clip,width=1.5\columnwidth]{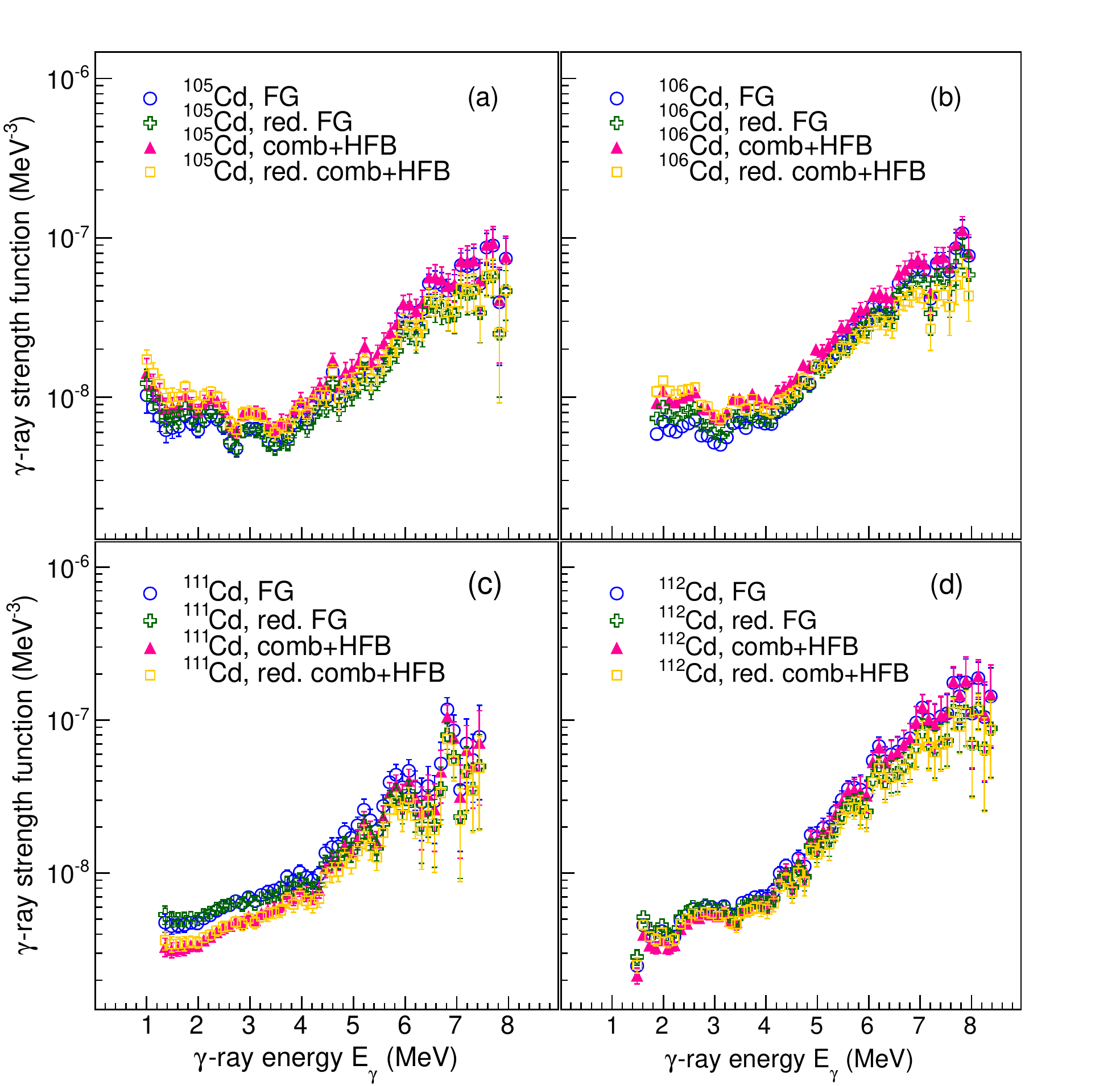}
 \caption {(Color online) Gamma-ray strength functions 
 of (a) $^{105}$Cd, (b) $^{106}$Cd, (c) $^{111}$Cd, and (d) $^{112}$Cd for the four
 different normalization approaches on the level densities. }
 \label{fig:rsfallnorm}
 \end{center}
 \end{figure*}

The normalized $\gamma$ strength functions for the four different level-density normalizations
of $^{105,106,111,112}$Cd are shown in Fig.~\ref{fig:rsfallnorm}. 
%---------------------------------------------------%
%---------------------------------------------------%
We clearly see a 
difference in the strength for $E_\gamma < 4$ MeV for the heavier $^{111,112}$Cd compared to the lighter 
$^{105,106}$Cd. For the latter, the tendency is a more flat and even a slightly increasing 
$\gamma$-strength function, while for the
former the $\gamma$ strength is decreasing when $E_\gamma$ decreases. Although there is no strong
low-energy enhancement as in Fe or Mo, it could indicate that this is the transitional mass region  
for the low-energy enhancement of the $\gamma$ strength. 

Another observation is that all the Cd strength functions seem to change slope at $E_\gamma \approx 4$ MeV. 
Above this value, the slope is significantly steeper than for lower $\gamma$ energies. This has
previously been seen in Sn isotopes~\cite{Sn_RSF,Heidi_Sn2}. These issues will be further addressed 
in the following section.

\section{Comparison with other data and models}
\label{sec:dis}

As mentioned in the previous section, our Cd data on the $\gamma$-strength function 
lack a strong low-energy enhancement, although the lighter isotopes appear to have more low-energy 
strength than the heavier ones. 
In addition, it is very likely that some extra strength is present in the region of
$4 \leq E_\gamma \leq 8$ MeV. 

%---------------------------------------------------%
 \begin{figure}[bt]
 \begin{center}
 \includegraphics[clip,width=\columnwidth]{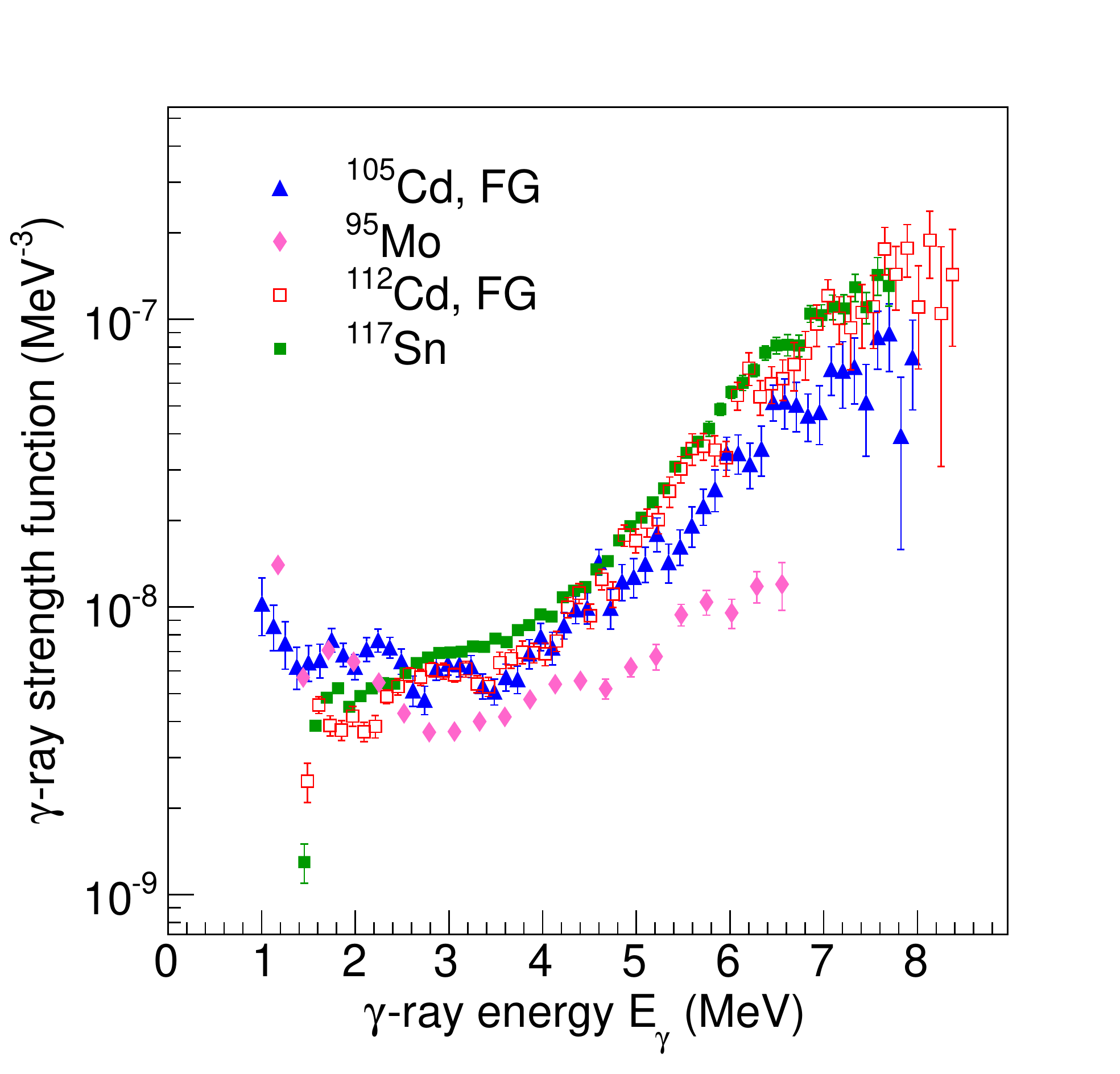}
 \caption {(Color online) Comparison of $\gamma$ strength functions of $^{95}$Mo, 
 $^{105,112}$Cd, and $^{117}$Sn (see text). }
 \label{fig:rsfcomp}
 \end{center}
 \end{figure}
%---------------------------------------------------%
In Fig.~\ref{fig:rsfcomp}, we have compared the strength functions of $^{105,112}$Cd
with $^{95}$Mo~\cite{Mo_RSF} and $^{117}$Sn~\cite{Sn_RSF}. It is very interesting to see 
how much $^{112}$Cd resembles $^{117}$Sn. On the other hand, $^{95}$Mo is very different from
both $^{112}$Cd and $^{117}$Sn, while $^{105}$Cd seems to be somewhat in between $^{95}$Mo
and $^{117}$Sn for $2 \leq E_\gamma \leq 4$ MeV. For higher $\gamma$ energies,
also $^{105}$Cd looks very much the same as $^{117}$Sn. 

To gain more insight of the observed $\gamma$ strength functions, 
we would like to compare our data with model calculations. One of the more widely used models 
for the $E1$ $\gamma$ strength is the Generalized Lorentzian 
(GLO) model~\cite{ko87,ko90}. This is a model tailored to give a reasonable description both on the
photoabsorption cross section in the GDR region, and on the $\gamma$ strength below the 
neutron separation energy. It is in principle dependent on the temperature of the final states $T_f$, 
which is in contradiction to the Brink hypothesis~\cite{brink}. However, by introducing a constant temperature,
the hypothesis is regained. 

The strength function within the GLO model is given by
\begin{align}
& f_{\rm GLO}(E_{\gamma},T_f) = \frac{1}{3\pi^2\hbar^2c^2}\sigma_{E1}\Gamma_{E1} \times \\ \nonumber
& \left[\frac{ E_{\gamma} \Gamma(E_{\gamma},T_f)}{(E_\gamma^2-E_{E1}^2)^2 + E_{\gamma}^2 
	\Gamma (E_{\gamma},T_f)^2} + \;0.7\frac{\Gamma(E_{\gamma}=0,T_f)}{E_{E1}^3} \right],
\label{eq:GLO}
\end{align} 
with
\begin{equation}
\Gamma(E_{\gamma},T_f) = \frac{\Gamma_{E1}}{E_{E1}^2} (E_{\gamma}^2 + 4\pi^2 T_f^2).
\end{equation}
The Lorentzian parameters $\Gamma_{E1}$, $E_{E1}$ and $\sigma_{E1}$ correspond to the width, 
centroid energy, and peak cross section of the GDR. We have made 
use of the parameterization 
of RIPL-2~\cite{RIPL} to estimate the GDR parameters as these are unknown experimentally for the 
individual Cd isotopes,
see Tab.~\ref{tab:rsfpar}. Because the even-even Cd isotopes are known to have a 
non-zero ground-state deformation~\cite{RIPL},
the GDR is split in two and we have therefore two sets of Lorentzian parameters (denoted by 
subscripts 1 and 2, see Tab.~\ref{tab:rsfpar}). 
%************************************************************************************%
\begin{table*}[htb]
\caption{Parameters used for the RSF models.} 
\begin{tabular}{lccccccccccccccc}
\hline
\hline
Nucleus    & $E_{E1,1}$ & $\sigma_{E1,1}$ & $\Gamma_{E1,1}$ & $E_{E1,2}$ & $\sigma_{E1,2}$ & $\Gamma_{E1,2}$ & $T_{\mathrm{min}}$ & $T_{\mathrm{max}}$ & $E_{M1}$ & $\sigma_{M1}$ & $\Gamma_{M1}$  & $E_{\mathrm{pyg}}$ & $\sigma_{\mathrm{pyg}}$ & $C_\mathrm{pyg}(T_\mathrm{min})$ & $C_\mathrm{pyg}(T_\mathrm{max})$\\
		   & (MeV)  	& (mb)            &  (MeV)        & (MeV)  	     & (mb)            &  (MeV)          & (MeV)       & (MeV)       & (MeV)    & (mb)          & (MeV)          & (MeV)              & (MeV)                   & ($10^{-7}$ MeV$^{-2}$)               & ($10^{-7}$ MeV$^{-2}$)   \\
\hline
$^{105}$Cd & 14.7       & 151.8           & 4.39          & 17.0         & 75.8            & 5.81            & 0.35        &  0.40       & 8.69     & 0.94          & 4.0            & 8.7(2)             & 1.5(1)                  & 2.2(2)                               &  1.1(1)     \\
$^{106}$Cd & 14.6       & 153.7           & 4.37          & 16.9         & 76.7            & 5.79            & 0.35        &  0.40       & 8.66     & 0.94          & 4.0            & 8.7(2)             & 1.5(1)                  & 2.4(2)                               &  1.1(2)     \\
$^{111}$Cd & 14.5       & 162.8           & 4.28          & 16.8         & 81.3            & 5.67            & 0.37        &  0.47       & 8.53     & 0.90          & 4.0            & 8.7(2)             & 1.5(1)                  & 2.9(3)                               &  1.7(2)     \\
$^{112}$Cd & 14.4       & 164.5           & 4.26          & 16.7         & 82.1            & 5.65            & 0.37        &  0.40       & 8.51     & 0.89          & 4.0            & 8.7(2)             & 1.5(1)                  & 3.7(3)                               &  2.4(4)     \\
\hline
\hline
\end{tabular}
\\
\label{tab:rsfpar}
\end{table*}
%************************************************************************************%
For the $M1$ strength, we have used a Lorentzian shape with the parameterization in Ref.~\cite{RIPL}. 

We treat the extra strength for high $\gamma$ energies in the same way as for 
the Sn isotopes~\cite{Sn_RSF,Heidi_Sn2}, adding a Gaussian-shaped pygmy resonance:
\begin{equation}
f_{\mathrm{pyg}} = C_{\mathrm{pyg}} \cdot\frac{1}{\sqrt{2\pi}\sigma_{\mathrm{pyg}}}
    \exp\left[\frac{-(E_{\gamma} - E_{\mathrm{pyg}})^{2}}{2\sigma_{\mathrm{pyg}}^2}\right].
\end{equation}
Here, $C_{\mathrm{pyg}}$ is a normalization constant, $\sigma_{\mathrm{pyg}}$ is the standard deviation, 
and $E_{\mathrm{pyg}}$ is the centroid of the resonance. 

The temperature of the final states is assumed to be constant, and is treated as a 
free parameter to get the best possible agreement with our data. As the normalization is uncertain, 
also the temperature is uncertain. In general, we get a slightly higher temperature for the normalization
options that give the largest low-energy $\gamma$ strength. We denote the temperature for the normalization
giving the largest low-energy strength $T_{\mathrm{max}}$, and the smallest low-energy strength $T_{\mathrm{min}}$.
The adopted $\gamma$-strength model parameters are given in Tab.~\ref{tab:rsfpar}.

As there are no photoneutron cross-section data on the individual Cd isotopes, 
we have compared our measurements with $(\gamma,x)$ data on natural Cd from Ref.~\cite{lepretreCd} 
and ($\gamma,n$) data on $^{106,108}$Pd taken from Ref.~\cite{hiro106Pd}. 
Assuming that the photoneutron cross section $\sigma_{\gamma}(E_\gamma)$ 
is dominated by dipole
transitions, we convert it into $\gamma$ strength by~\cite{RIPL}:
\begin{equation}
f_{\gamma }(E_\gamma) = \frac{1}{3\pi^2\hbar^2c^2}\frac{\sigma_{\gamma }}{E_\gamma}.
\end{equation}

\begin{figure}[bt]
 \begin{center}
 \includegraphics[clip,width=\columnwidth]{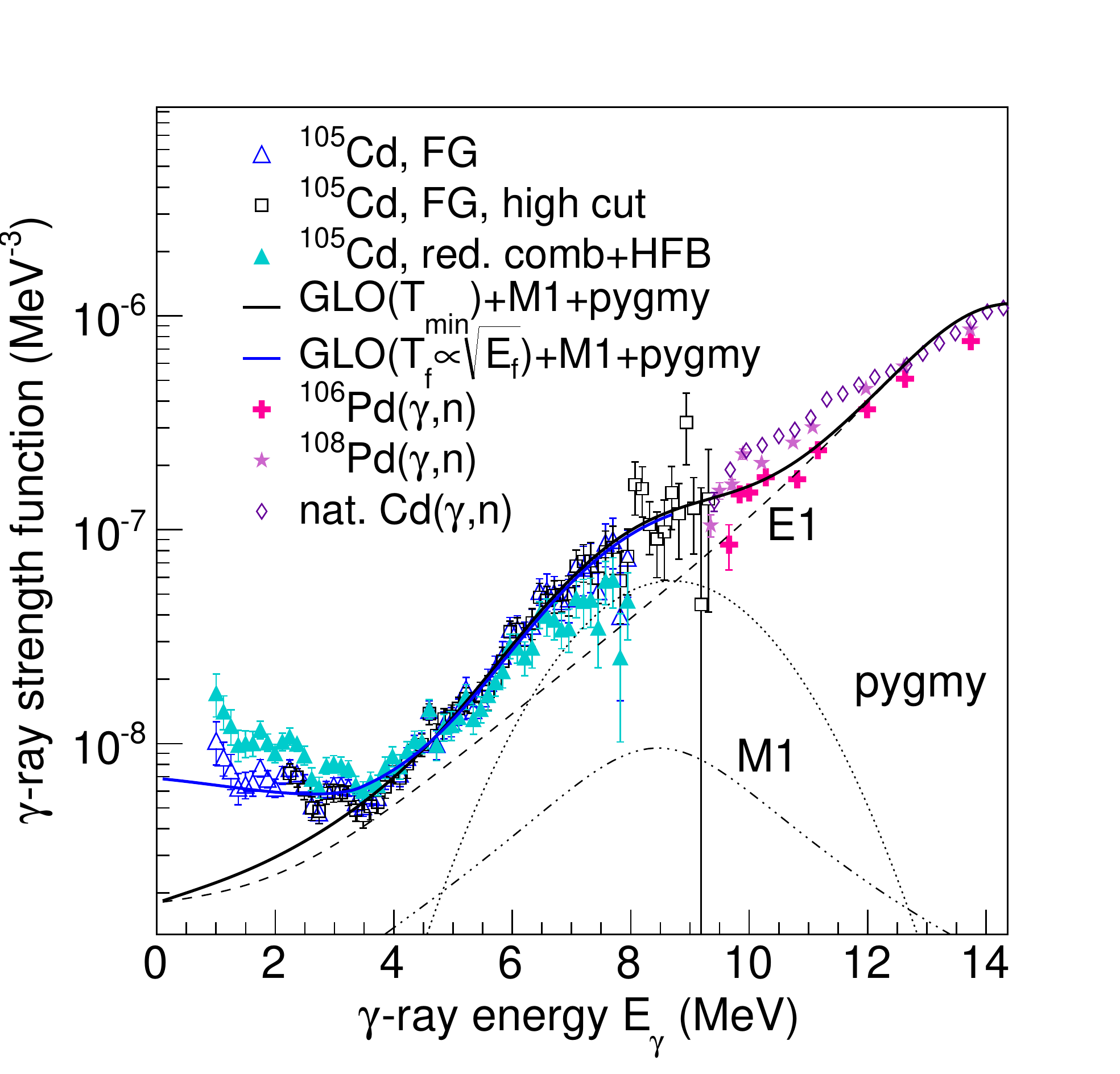}
 \caption {(Color online) Calculations using the GLO model with a constant temperature ($T_\mathrm{min}$) 
    and a variable temperature ($T_f \propto \sqrt{E_f}$) compared to data of $^{105}$Cd for the normalization
    giving the lowest possible low-energy strength (FG) and the highest (combinatorial-plus-HFB, reduced spin window).
    The black triangles show the extracted strength function for a higher cut on $E_\gamma$ and $E$ in the
    first-generation matrix of $^{105}$Cd. Photonuclear data from Refs.~\cite{lepretreCd,hiro106Pd} are also shown.}
 \label{fig:rsf105Cd}
 \end{center}
 \end{figure}

In Fig.~\ref{fig:rsf105Cd}, our data on the $\gamma$ strength function of $^{105}$Cd
and the photonuclear data are shown together with the model calculations for the lowest temperature
$T_\mathrm{min}$ in the GLO model. 
%---------------------------------------------------%
%---------------------------------------------------%
It can be seen that the calculations are in reasonable agreement with the Pd data
from Ref.~\cite{hiro106Pd} and our data down to $E_\gamma \approx 3.5$ MeV. For lower $\gamma$ energies,
our data show significantly more strength than the constant-temperature calculations. 

Because $\gamma$ decay has a considerable probability also above $S_n$ for $^{105}$Cd, 
see Fig.~\ref{fig:alfnaun}, we have extracted the strength function for this nucleus up to 
$E_\gamma \approx 9.3$ MeV. This is done by choosing a higher $E_\gamma$ limit of 2.25 MeV in the 
first-generation matrix to ensure
that we do not mix with data from the $^{104}$Cd channel. The resulting strength function is displayed 
in Fig.~\ref{fig:rsf105Cd} as open squares. Although the statistical errors are quite large, 
we are able to bridge the gap up to the
$(\gamma,n)$ measurements, thus further supporting the presence of an enhanced strength
in the $6-10$ MeV region. It is also a strong indication that the $\left< \Gamma_\gamma \right>$ value we have
chosen for normalization is reasonable.  

\begin{figure*}[bt]
 \begin{center}
 \includegraphics[clip,width=1.7\columnwidth]{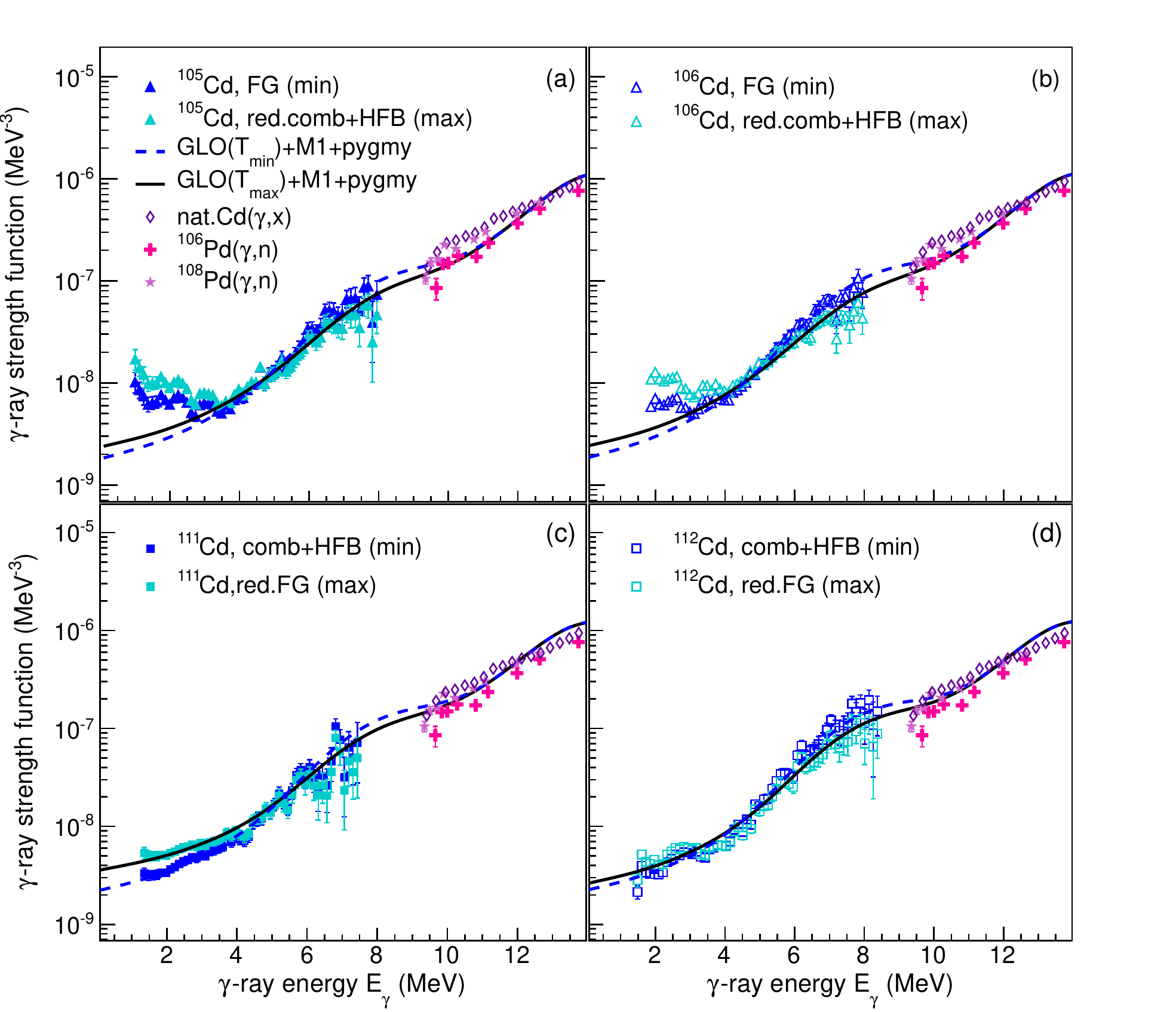}
 \caption {(Color online) Modeled $\gamma$ strength functions compared to the data for (a) $^{105}$Cd,
    (b) $^{106}$Cd, (c) $^{111}$Cd, and (d) $^{112}$Cd for the level-density normalizations which give 
    the minimum or maximum strength at low $\gamma$ energies. Photonuclear data from 
    Refs.~\cite{lepretreCd,hiro106Pd} are also displayed.}
 \label{fig:allmod}
 \end{center}
 \end{figure*}

The resulting $\gamma$-strength models for all the Cd isotopes studied here are shown together 
with our data and the photonuclear
data in Fig.~\ref{fig:allmod}.
%---------------------------------------------------%
%---------------------------------------------------%
We observe that the models fit our data quite well, in particular for $^{111,112}$Cd.
The extra strength between $E_\gamma \approx 5-8$ MeV seems to be well described by a Gaussian
function just as in the Sn case. 

As of today, the origin of this strength is not well understood. It could 
be due to enhanced probability for $E1$ transitions due to the so-called neutron skin oscillation, 
see Refs.~\cite{Sn_RSF,Heidi_Sn2} and references therein. There is also a possibility that the 
$M1$ spin-flip resonance gives more strength than the parameterization we have adopted here. 
In a recent work on $^{90}$Zr by Iwamoto \textit{et al.}~\cite{iwa90Zr}, it is shown how both
an $E1$ pygmy dipole resonance and an $M1$ resonance are present in the energy region
$E_\gamma \approx 6-11$ MeV, with similar strengths. It could be that the same is the case 
also for the Cd isotopes. Unfortunately, with our experimental technique it is not possible 
to separate $E1$ and $M1$ transitions in the $\gamma$ strength.
It would therefore be highly desirable to investigate this further with the
experimental technique applied in Ref.~\cite{iwa90Zr}.

Assuming that all the pygmy strength is of $E1$ type, we have compared the energy-integrated strength
of this structure with the classical energy-weighted Thomas-Reiche-Kuhn (TRK) sum rule (without exchange forces) 
given by~\cite{TRK}:
\begin{equation}
\sigma_\mathrm{TRK} \simeq 60 \frac{NZ}{A}\:\left[{\mathrm{MeV\cdot mb}}\right].
\end{equation}
The results are shown in Tab.~\ref{tab:trk}. 
%************************************************************************************%
\begin{table}[htb]
\caption{Maximum and minimum integrated strengths of the pygmy resonance.} 
\begin{tabular}{lcccc}
\hline
\hline
Nucleus    & $\sigma(T_{\mathrm{max}})$ & $\sigma(T_{\mathrm{min}})$ & TRK       & \% of TRK  \\
		   & (MeV mb)  	    & (MeV mb)       &  (MeV mb) &       \\
\hline
$^{105}$Cd & 11.3           & 21.8           & 1563.4    & $0.7-1.4$      \\
$^{106}$Cd & 11.3           & 24.4           & 1575.9    & $0.7-1.5$      \\
$^{111}$Cd & 17.4           & 28.7           & 1634.6    & $1.1-1.8$      \\
$^{112}$Cd & 24.4           & 37.4           & 1645.7    & $1.5-2.3$      \\
\hline
\hline
\end{tabular}
\\
\label{tab:trk}
\end{table}
%************************************************************************************%

The uncertainty of the normalization gives a rather large uncertainty in the fraction of the 
sum rule, but the general trend is an increasing pygmy strength as the neutron number increases. 
This is in agreement with expectations based on the neutron-skin oscillation mode, see for example 
Ref.~\cite{daoutidis}.

We note that for $^{105,106}$Cd, the models underestimate the strength for $E_\gamma < 3$ MeV. 
Also, we find it not possible to compensate for this by just increasing $T_f$, because then 
the overall strength will be too large for the data at higher $\gamma$ energies. 
In an attempt to describe the extra strength at low $\gamma$ energies, we have tested a variable 
temperature of the final levels, $T_f \propto \sqrt{E_f}$, in the GLO model. 
This is shown as a solid, blue line in Fig.~\ref{fig:rsf105Cd}. It is seen that the variable-temperature 
model is rather successful in describing the low-energy data for the normalization giving the lowest
low-energy strength.

It is however hard to explain why one should have a constant temperature for $^{111,112}$Cd and
a variable one for $^{105,106}$Cd. By inspecting the level densities, they all have an approximately
constant slope in log scale, compatible with a constant-temperature level density 
$\rho_\mathrm{CT}(E) \propto \exp(E/T)$. This has recently been supported by particle-evaporation 
experiments in lighter nuclei~\cite{alex_rapid}. In addition, the variable-temperature approach is 
not able to reproduce the data normalized to give maximum strength at low $\gamma$ energies
(reduced spin range for the initial levels). We therefore conclude that it is more probable that
some low-lying strength is present below $E_\gamma \approx 3.5$ MeV for $^{105,106}$Cd, similar as
for the Mo isotopes but not as strong. However, one must keep in mind that 
the uncertainty in the level-density normalization hampers
any firm statements. Further studies
of nuclei in this mass region are ongoing, and will hopefully shed more light on this issue.

\section{Summary and outlook}
\label{sec:sum}
The level densities and $\gamma$-ray strength functions of $^{105,106,111,112}$Cd have been deduced from
particle-$\gamma$ coincidence data using the Oslo method. The level densities are in excellent agreement 
with known levels at low excitation energy. We note that the slope in level density decreases from the 
heavier $^{111,112}$Cd to the lighter $^{105,106}$Cd. This is probably due to the neutron number approaching
the $N=50$ closed shell.

The $\gamma$-ray strength functions for all the Cd isotopes display an enhancement for $E_\gamma > 4$ MeV, 
very similar to features observed in the previously studied Sn isotopes. The nature of this extra 
strength could
not be determined in the present work, but could in principle be due to both $E1$ and $M1$ transitions. 
Future investigations are highly desirable to resolve these multi-polarities.Ê

At $\gamma$-ray energies below 3 MeV, the $\gamma$-strength function of the lighter $^{105,106}$Cd 
isotopes show an increase compared to $^{111,112}$Cd. Although this might be due to the vicinity of the 
$N=50$ shell closure and the resulting reduced level density in the lighter isotopes, it is more likely that 
this work uncovered the mass region exhibiting the onset of the low-energy enhancement. 
Further measurements are in progress and the results will provide more details regarding this transitional region.

\begin{acknowledgments}
We are very grateful to C.~Scholey and the nuclear physics group at the University of Jyv\"{a}skyl\"{a} 
(JYFL) for 
lending us the $^{106,112}$Cd targets.
Funding of this research from the Research Council of Norway, project grants no. 180663 and 205528, 
is gratefully acknowledged. 
M. W. acknowledges support from the National Research Foundation of South Africa.
We would like to give special thanks to 
E.~A.~Olsen, A.~Semchenkov, and J.~Wikne for providing the beam. 
\end{acknowledgments}

%\vfill
\end{document}